# Theory of extreme optical concentration in all-dielectric waveguides

Nazmus Sakib[1] and Judson D. Ryckman[1,*]

[1]*Clemson University, Holcombe Department of Electrical and Computer Engineering*
*Clemson, SC, 29634 USA*

We introduce transversely structured all-dielectric waveguides which exploit the vectorial nature of light to achieve extreme sub-wavelength confinement in high index dielectrics, enabling characteristic mode dimensions below $\lambda_0^2/1,000$ without metals or plasmonics. We also derive the metric of 'optical concentration' and show its convenient usage in characterizing enhanced linear and non-linear interactions at the nanoscale. This work expands the 'toolbox' of nanophotonics and opens the door to new types of ultra-efficient and record performing linear and nonlinear devices with broad applications spanning classical and quantum optics.

## I. INTRODUCTION

The waveguide is perhaps one of the most important and versatile building blocks utilized in electromagnetics and modern nanophotonics, where it is heavily relied on to enable advancements in opto-electronics [1], opto-mechanics [2], non-linear optics [3–5], quantum photonics [6], nano-manipulation [7], and nano-sensing [8,9]. Waveguides capable of sub-wavelength light confinement are particularly advantageous for realizing ultra-efficient active photonic components such as classical and quantum light sources [10–14], phase/amplitude modulators [15–18], photodetectors [19], and atom-light interfaces [20].

Over the past two decades a large dichotomy between plasmonic and dielectric based photonics has emerged. To date, nanoplasmonic waveguides integrating metals are unrivalled in terms of achieving ultra-small mode dimensions [21–24]. However, ohmic losses result in significant passive propagation losses (~dB μm$^{-1}$) at optical frequencies which are untenable for many applications. Dielectric waveguides on the other hand, offer significantly lower passive propagation losses (~dB cm$^{-1}$ to dB m$^{-1}$), yet are often implemented in resonant or slow-light structures in an effort to enhance temporal interaction and derive a larger response per unit energy. Dielectric resonators and band edge devices, however, are restricted to operating in narrow optical bandwidths and may require active resonant tuning to stabilize amidst environmental variations – either of these factors may be prohibitive in certain applications.

In this article, we theoretically explore an alternative waveguide architecture, capable of supporting enhanced nanoscale light concentration without the losses and bandwidth limitations of existing approaches. As recent all-dielectric metamaterial investigations to this same problem have recently indicated [25], we find that a promising yet largely unexplored regime exists in the case of high-index contrast media structured on the subwavelength scale. This regime specifically exploits the vectorial nature of light and enables the design of all-dielectric waveguides featuring extreme optical concentrations – a metric which is herein derived and related to both linear and non-linear devices. Implications of such designs are considered and include: record low optical mode areas and high Purcell factors for all-dielectric waveguides; ultra-low active volume (high index, e.g. solid-state) linear components; and, perhaps surprisingly, the simultaneously ability to either suppress or enhance core non-linearity.

## II. APPROACH

In an inhomogeneous medium the mode solutions to Maxwell's equations are constrained by the presence of vectorial boundary conditions which must be enforced at all interfaces. Specifically, optical fields exhibit discontinuities in the normal component of $\mathcal{E}$ and tangential component of $\mathcal{D}$ at any interface with dielectric contrast ($\epsilon_h > \epsilon_l$) owing to the two boundary conditions, summarized here for non-dispersive dielectrics:

$$\mathcal{D}_{\perp,l} = \mathcal{D}_{\perp,h} \rightarrow \mathcal{E}_{\perp,l} = \frac{\epsilon_h}{\epsilon_l}\mathcal{E}_{\perp,h} \qquad (1)$$

$$\mathcal{E}_{\parallel,h} = \mathcal{E}_{\parallel,l} \rightarrow \mathcal{D}_{\parallel,h} = \frac{\epsilon_h}{\epsilon_l}\mathcal{D}_{\parallel,l} \qquad (2)$$

Fig. 1 illustrates simple configurations where these boundary conditions can enhance sub-wavelength optical concentration by locally enhancing the electric field energy density $u_E = \frac{1}{2}\boldsymbol{D} \cdot \boldsymbol{E}$. The first boundary condition, Eq. (1), requires continuity of the normal component of electric displacement and famously yields enhancement of the electric field in the low index medium, which is widely exploited in plasmonic and slot waveguides, e.g. Fig. 1(ii), under the appropriate field polarization [26–28]. While this 'slot effect' has proven to be especially useful for sensors [9], nanomanipulation in aqueous media [7], and low index integrated (e.g. polymer) linear and non-linear devices [18], it is not a particularly useful configuration for developing solid-state components based on high index materials such as lasers [29], quantum emitters [30], phase/amplitude modulators [1,31], or photodetectors [19].

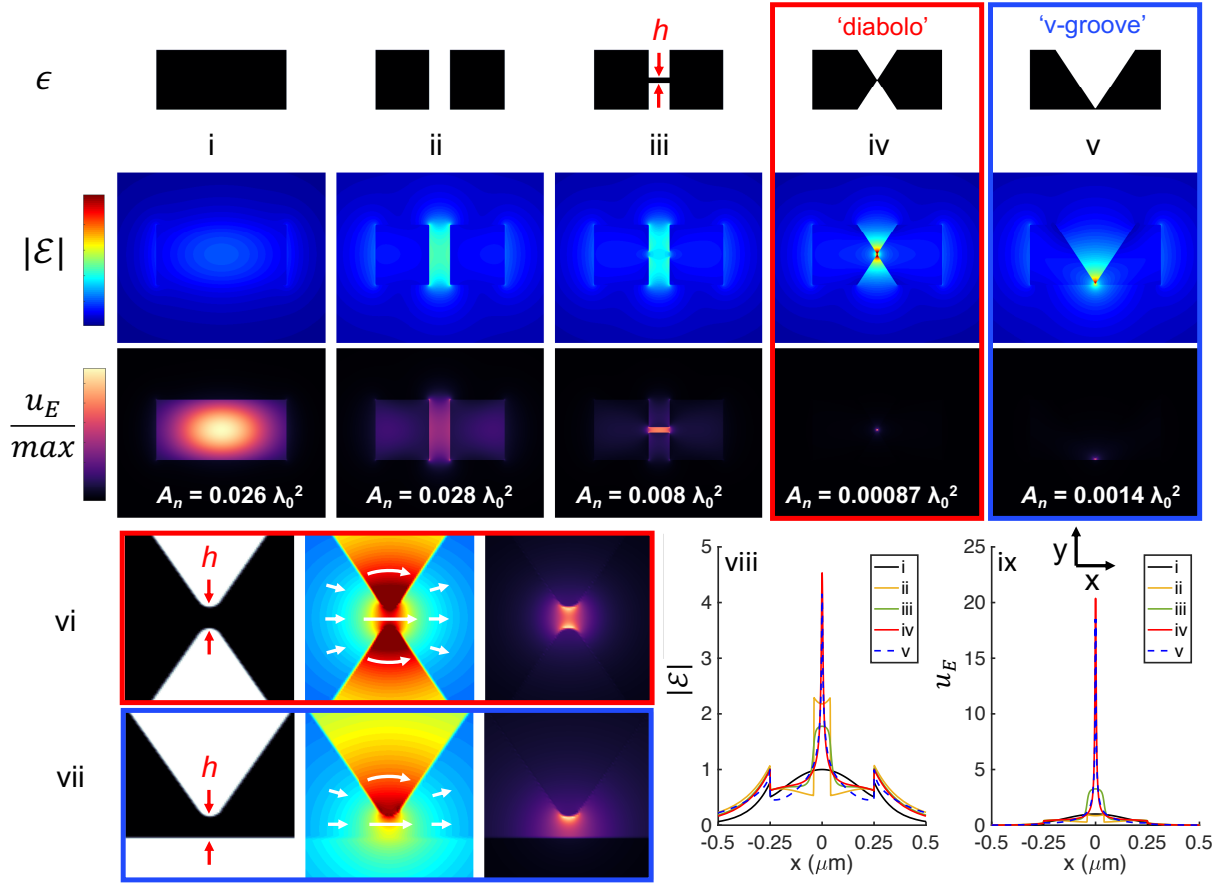

**Fig. 1.** Dielectric function, electric field (energy normalized, single color scale), and electric field energy density (peak normalized, individual color scales) for Si/SiO$_2$ waveguide quasi-TE fundamental modes simulated at $\lambda_0 = 1.55\ \mu m$: (i) strip (500 nm x 220 nm), (ii) slot (80 nm), (iii) bridged slot (80 nm, $h$ = 20 nm), (iv) diabolo ($h$ = 5 nm), and (v) v-groove ($h$ = 5 nm) waveguides. (vi) and (vii) reveal a zoomed 40 nm x 40 nm view of (iv) and (v), including illustration of the local field vectors. (viii) Cross section of electric field and (ix) energy density (energy normalized where peak value of A is set to unity).

However, the second boundary condition Eq. (2), which has recently been highlighted and exploited in the design of ultra-low mode volume photonic crystal cavities [5,32,33], reveals that the electric displacement, and thus $u_E$, can be locally enhanced within a *high index* medium. If a narrow ($h$ = 20 nm) high index bridge is introduced to span the slot ($w$ = 80 nm) of a conventional slot waveguide, as shown in Fig. 1(iii), the predominant and already 'slot enhanced' $\mathcal{E}_x$ field component of the strongly polarized quasi-TE mode is carried through the high index bridge owing to continuity of the tangential component of the electric field. Such a configuration has a compound effect on the electric field energy density and effectively squares the energy density enhancement provided by the slot effect while enabling the peak optical energy density to carry into the high index medium. As a result, the local electric field energy density can be enhanced up to a total factor of approximately $(\epsilon_h/\epsilon_l)^2 = (n_h/n_l)^4$ relative to a homogenous waveguide core. This corresponds to a potential energy density enhancement (and mode area suppression) factor of ~30 for an oxide cladded silicon structure or ~150 for an air cladded device.

Designs which improve upon the bridged slot geometry of Fig. 1(iii) and better approach the $\sim(n_h/n_l)^4$ enhancement factor limit may be achieved by tailoring the structure to enforce boundary Eq. (2) only where the slot effect is maximized, i.e. by using a diabolo or v-groove geometry as shown in Figs. 1(iv) and 1(v). These diabolo and v-groove designs function similar to plasmonic bow-ties and v-grooves [23], except they uniquely foster strong optical concentration in high index materials while providing the inherent advantages of low-loss all-dielectric media. For prototypical silicon/air and silicon/SiO$_2$ dielectrics considered here, we observe record low waveguide mode areas, $A_n \sim \lambda_0^2/1{,}000$ to $\sim \lambda_0^2/10{,}000$, and high energy densities. The enhanced characteristics are observed to be a particularly strong function of the index contrast and high index 'bridge' height $h$ (Fig. 3); and a weak function of the groove tip's nanoscaled radius of curvature $r$ (Fig. S1).

To assess the characteristics and implications of a waveguide with extreme electric field energy density enhancement, with generality, we are motivated to derive and summarize the concept of "optical concentration" in the context of optical waveguides. As discussed by Miller [34], optical concentration is a unifying yet underutilized metric which fundamentally limits the performance scaling in any active photonic device. The introduction of this concept importantly offers a simple framework (and intuitive alternative to invoking 'local density of states'), which unifies the disparate and sometimes limited metrics of: *i*) 'classic' mode area $A_n$, *ii*) Purcell factor $F_P$, *iii*) confinement factor $\Gamma$, and iv) non-linear effective mode area $A_{eff}^{(NL)}$. For example, $A_n$ and $F_P$ enable one to quantify the enhancement in the spontaneous emission rate for a dipole (atomic scale volume) in a resonant mode field [35,36], yet they do not easily map to the characterization of different sorts of modern active photonic devices with arbitrary active volume dimensions. Similarly, optimization of $\Gamma$ does not result in global minimization of active volume, but rather is strictly limited to minimization of device length $l$.

### III. OPTICAL CONCENTRATION

Optical concentration is closely related to the electromagnetic energy density, $u$, which is defined as [37]:

$$u = \frac{1}{2}[\boldsymbol{D} \cdot \boldsymbol{E} + \boldsymbol{H} \cdot \boldsymbol{B}] \quad (3)$$

where the first term ½$\boldsymbol{D} \cdot \boldsymbol{E}$ describes the electric field energy density $u_E$ stored in a medium, including that in the propagating electric field and electric polarization, expressed here for local isotropic dielectric media as:

$$u_E = \frac{1}{2}\frac{\partial(\omega\varepsilon(\boldsymbol{r},\omega))}{\partial \omega}|\boldsymbol{E}(\boldsymbol{r},\omega,t)|^2 \approx \frac{1}{2}\varepsilon(\boldsymbol{r})|\boldsymbol{E}(\boldsymbol{r},t)|^2 \quad (4)$$

which simplifies to the right most expression in the approximation of minimally dispersive dielectric materials where $\varepsilon(\boldsymbol{r}) = \varepsilon_0 \varepsilon_r(\boldsymbol{r})$ is the permittivity profile of the structure.

To quantify the important nature of the electric field energy density $u_E$ in waveguides, and thus the resulting optical concentration, we reformulate the classic variational method applied to non-leaky waveguides [38], in terms of the electromagnetic energy density and a time average perturbation $\langle \Delta u \rangle$:

$$\Delta\tilde{\beta} = \frac{\omega \int \langle \Delta u \rangle dA}{\omega \int \langle (\partial u / \partial |\boldsymbol{k}|) \cdot \hat{z}\rangle dA} \quad (5)$$

This expression could be interpreted to quantify a complex change in power per unit length, normalized by the time averaged total energy flux across a plane perpendicular to the z-axis (unit power) [39], and should describe the complex phase shift of the wave in the z-direction per unit length. It has been shown that this type of variational approach leads directly to a rigorous derivation of the optical confinement factor $\Gamma_\mathcal{A}$ [40] (see Appendix 3) which satisfies:

$$\Delta\tilde{\beta} = \frac{\omega}{c}\Gamma_\mathcal{A}\Delta\tilde{n}_\mathcal{A} \quad (6)$$

where $\Delta\tilde{n}_\mathcal{A}$ is a complex index perturbation uniformly applied to an active area $\mathcal{A}$ in the waveguide cross-section.

Let us now consider an active waveguide device, with uniform cross-section and active area $\mathcal{A}$, which is extended along a propagation length $l$ such that the total active volume is $\mathcal{V} = l\mathcal{A}$, as illustrated in Fig. 2. The total accumulated complex phase response of a linear active device is proportional to the confinement factor times the device length:

$$\Delta\tilde{\beta}l \propto \Gamma_\mathcal{A} l \quad (7)$$

From this commonly utilized expression however, it is unclear how the device response scales or depends on the active volume $\mathcal{V}$. Given that the dimensions of active volume are a critical factor in real devices, for example in dictating the minimum energy consumption scaling of solid-state devices where energy can potentially be locally delivered to deeply sub-wavelength areas and volumes [34], it would be valuable to instead quantify the accumulated response $\Delta\tilde{\beta}l$ in terms of active volume $\mathcal{V}$.

Thus, we introduce a definition of optical concentration $U_\mathcal{A}$ [m$^{-2}$], consistent with Eq. (4) and (5) as:

$$U_\mathcal{A} \equiv \frac{\int_\mathcal{A} \langle u_E \rangle dA}{\mathcal{A}}\left(\frac{c}{n_\mathcal{A}}\right)\frac{2}{\omega \int \langle (\partial u / \partial |\boldsymbol{k}|) \cdot \hat{z}\rangle dA} \quad (8)$$

Which can alternatively be expressed (see Appendix 3) in relation to the rigorously defined confinement factor from Eq. (6):

$$U_\mathcal{A} = \frac{\Gamma_\mathcal{A}}{\mathcal{A}} = \eta \frac{1}{\mathcal{A}} \frac{\int_\mathcal{A} u_E dA}{\int u_E dA} = \eta \frac{\gamma_\mathcal{A}}{\mathcal{A}} \quad (9)$$

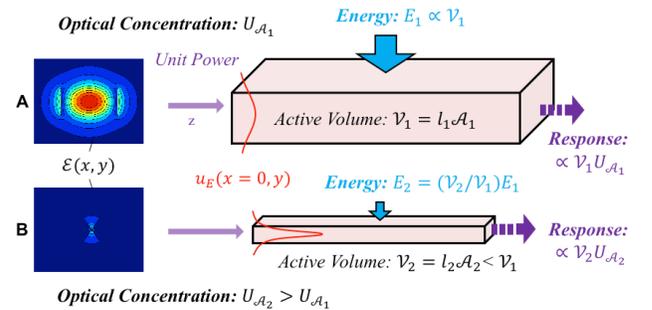

**Fig. 2.** Efficiency and active volume scaling principles. Under a constant desired linear-optic response, active volume is minimized when optical concentration $U_\mathcal{A}$ is maximized.

This newly defined metric of optical concentration captures both the effects of: i) longitudinal concentration via a factor $\eta = \frac{n_g}{n_\mathcal{A}}$, where $n_g$ is the waveguide group index; and ii) transverse concentration via a factor $\frac{\gamma_\mathcal{A}}{\mathcal{A}}$, which computes the average electric field energy $u_E$ in the active area per unit length, normalized to the total electric field energy per unit length. Thus, it retains a clear dependence on $u_E$, which can be locally enhanced roughly two orders of magnitude in our diabolo and v-groove waveguide designs.

With the optical concentration so-defined, the accumulated optical response originating from linear matter-light interaction now depends exactly on the product of the optical concentration and active volume:

$$\Delta\tilde{\beta} l = \frac{\omega}{c} U_\mathcal{A} \mathcal{V} \Delta\tilde{n}_\mathcal{A} \propto U_\mathcal{A} \mathcal{V} \qquad (10)$$

The optical concentration is therefore the coefficient which satisfies the relations $\Delta n_{eff} l = \Delta n_\mathcal{A} U_\mathcal{A} \mathcal{V}$, $g_n l = g_\mathcal{A} U_\mathcal{A} \mathcal{V}$, and $\alpha_n l = \alpha_\mathcal{A} U_\mathcal{A} \mathcal{V}$, where $\Delta n_{eff}$ is the perturbation in waveguide effective index arising from a perturbation $\Delta n_\mathcal{A}$ of the active region with active volume $\mathcal{V}$, and $g_n$ or $\alpha_n$ are the modal gain or absorption coefficients imparted onto the waveguide mode from the gain or absorption coefficients $g_\mathcal{A}$ or $\alpha_\mathcal{A}$ of the active medium.

Unlike Eq. (7), the expression in Eq. (10) highlights a scaling principle which serves as a cornerstone of modern nanophotonics [34]. For a given material platform, reducing energy consumption in active devices requires simultaneous reduction in physical active volume $\mathcal{V}$ *and* enhancement of optical concentration $U_\mathcal{A}$, as illustrated in Fig. 2. Thus, the linear waveguide device operating with the smallest active volume does not necessarily feature the smallest device length and largest confinement factor, but rather exhibits the highest optical concentration $U_\mathcal{A}$.

With this framework in place, we can also consider more generally the problem of achieving a large waveguide optical concentration $U_\mathcal{A}$. A preferred solution would exhibit the following traits: (1) all-dielectric design to avoid the loss limitations associated with metals and plasmonics, (2) potential for broadband operation without relying on resonance or band edge effects, (3) achieves optical concentration in a high index material (e.g. semiconductor) to facilitate solid-state active components, and (4) potential compatibility with planar integrated photonics. While items (1-4) are fostered with our designs, as we will show, another elegant approach to this problem is resonance-free light recycling [41,42]. Using a mode division multiplexing strategy this solution achieves a waveguide optical concentration which is the linear sum of the modal concentrations $U_\mathcal{A} = \sum_{n=1}^{N} U_{\mathcal{A},n}$ for each of the $N$ forward and backward propagating modes superimposed in the structure. Assuming this strategy could be applied to ~6 spatial modes on each of 2 polarizations (e.g. $N = 12$), the waveguide concentration factor can be increased by approximately an order of magnitude while retaining a roughly constant active volume $\mathcal{V}$. However, this approach increases the accumulated waveguide loss by a factor on the order of ~$N$ and in general may be restricted to configurations where the active area $\mathcal{A}$ is comparable to the diffraction limit ~$(\lambda_n/2)^2$. In the waveguide designs under consideration here, we are exploring an alternative regime which achieves extreme optical concentrations $U_\mathcal{A}$ for sub-diffraction active areas, $\mathcal{A} < (\lambda_n/2)^2$.

For a given waveguide mode, the optical concentration $U_\mathcal{A}$ from Eqs. (8) or (9) is a strong function of both the particular placement and geometry of the active area $\mathcal{A}$ and the mode's energy density distribution $u_E$. Most photonic devices will employ a finite non-zero active area $\mathcal{A}$. In general terms however, the maximum theoretical optical concentration occurs in the case of an infinitesimal active area $\mathcal{A} \to dA$ centered at the location of maximum energy density $\boldsymbol{r}_{max}$, such that:

$$U_{\mathcal{A} \to dA}|_{r=r_{max}} = \frac{\eta}{A_n} \qquad (11)$$

which recovers the appropriate definition of the 'classic' mode area $A_n$ in non-leaky waveguides, applicable toward the calculation of the waveguide Purcell factor and in determining the enhancement of spontaneous emission for an emitter placed at $\boldsymbol{r}_{max}$ [36], where the mode area becomes:

$$A_n = \frac{\int u_{E_n} dA}{u_{E_n}(\boldsymbol{r}_{max})} \qquad (12)$$

The Purcell factor in a waveguide is thus related directly to the optical concentration according to:

$$F_P = \left(\frac{3}{2\pi}\right) \lambda_n^2 U_\mathcal{A} \to \frac{3}{2\pi}\left(\frac{\eta}{\tilde{A}_n}\right) \qquad (13)$$

where $\lambda_n = \lambda_0/n_\mathcal{A}$ and $\tilde{A}_n = A_n(\lambda_n)^{-2}$ is the normalized mode area in units of $\lambda_n^2$. In a broadband waveguide the longitudinal concentration factor remains $\eta = \frac{n_g}{n_\mathcal{A}}$. However, if the waveguide is formed into a standing wave cavity mode with finite finesse $\mathfrak{F}$, the on-resonance Purcell factor can be calculated by replacing the longitudinal factor $\eta$ from Eq. (13) with $\eta = \mathfrak{F}/\pi$ to recover the famous $F_P \propto Q/V$ form [34–36]. The mode area $A_n$ in our diabolo or v-groove waveguide, should therefore be suppressed, and $F_P$ correspondingly enhanced, relative to a homogenous core waveguide by a factor approaching the maximum energy density enhancement factor ~$(n_h/n_l)^4$.

## IV. RESULTS / DISCUSSION

Fig. 3 reports the modal characteristics of air and oxide cladded silicon diabolo and v-groove waveguides (e.g. from Fig. 1) as a function of the silicon bridge height $h$. In this analysis, the v-groove bottom cladding is fixed to oxide whereas a symmetrically distributed cladding material is considered for the diabolo geometry (see Appendix: Methods for additional detail). As shown in Fig. 3(a), the mode area $A_n$ of both waveguides decreases significantly with decreasing $h$, reaching values in the range $A_n \sim \lambda_0^2/1{,}000$ to $\sim \lambda_0^2/10{,}000$, more than one to two orders of magnitude below the diffraction limit for bulk silicon. The diabolo geometry is observed to enable the smallest values of $A_n$, which is attributed to the centered placement of the bridge and corresponding mode symmetry. Unlike slot waveguides which achieve $\sim(n_h/n_l)^2$ enhanced optical concentration solely in a low index medium, the v-groove and diabolo waveguides offer $\sim(n_h/n_l)^4$ enhancement in optical concentration (and $1/A_n$, $F_P$) in a *high index* medium. Also unlike a slot waveguide, the nanoscale bridge dimensions of these structures are expected to be compatible with the critical dimensions of standard photolithography (i.e. >150-300 nm) since the grooves can be realized by anisotropic wet etching of crystalline silicon [43,44].

Fig. 3(b) reports the non-linear (NL) effective mode area $A_{eff}^{(NL)}$ for both waveguides, which is observed to exhibit substantially different characteristics and trends with respect to refractive index contrast and waveguide geometry than the 'classic' mode area. In both structures evaluated in oxide claddings, reduction of $h$ results in increasing values of $A_{eff}^{(NL)}$, indicative of non-linearity *suppression*. Meanwhile in the air clad diabolo waveguide, record low values of $A_{eff}^{(NL)}$, smaller than any existing silicon nanowire geometry [45], not employing slow-light effects [46], are predicted.

Here, the non-linear mode area calculation assumes the non-linearity arises strictly from the core material (e.g. silicon) in the approximation of single-mode degenerate four-wave mixing (FWM). Unlike linear 'matter-light' interaction metrics, which we've shown to be proportional or inversely proportional to optical concentration $U_{\mathcal{A}}$, the non-linear mode area $A_{eff}^{(NL)}$ captures a distinctly different phenomenon of 'light-matter-light' interaction. An accurate description of $A_{eff}^{(NL)}$ in high-index inhomogenous media, is known to require a fully vectorial approach which accounts for the exact near-field distribution and group velocity [47,48]. In the literature however, there is generally no clear linkage between $A_{eff}^{(NL)}$ and other metrics used to characterize linear 'matter-light' interactions. We've recently derived such a linkage (see Appendix 5) and present an alternative formula for $A_{eff}^{(NL)}$, which is both rigorous and intuitive, and agrees with other fully vectorial reports [47,48]:

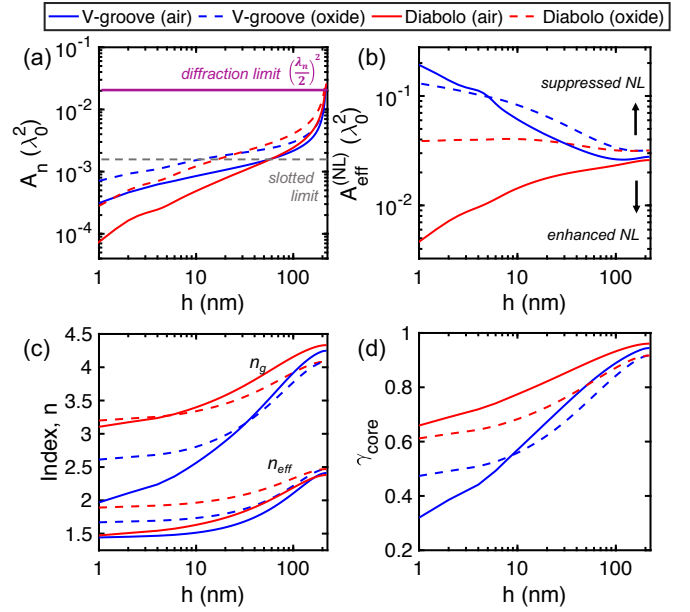

**Fig. 3.** All dielectric v-groove and diabolo waveguide modal characteristics as a function of bridge height $h$: (a) 'classic' mode area, (b) non-linear effective mode area, (c) group and effective indices, and (d) core medium confinement factor. Note: $h = 220$ nm corresponds to an unmodified strip waveguide.

$$A_{eff}^{(NL)} = \frac{1}{(U_{\mathcal{A}} \mathcal{A})^2} \frac{\left(\iint_{\mathcal{A}} |\mathcal{E}|^2 dA\right)^2}{\iint_{\mathcal{A}} |\mathcal{E}|^4 dA} \qquad (14)$$

where the term $U_{\mathcal{A}} \mathcal{A} = \Gamma_{\mathcal{A}}$. Unlike linear metrics (e.g. $A_n, F_P, \Gamma_{\mathcal{A}}$), the non-linear effective mode area $A_{eff}^{(NL)}$ depends on the square of optical concentration and active area $\mathcal{A}$, with an additional corrective term that factors in the $|E|^4$ profile rather than simply the $u_E$ profile.

For the core nonlinearity considered in Fig. 3(b), the active area $\mathcal{A}$ consists of the entire high index portion of the waveguide, e.g. $'core' = \mathcal{A} \rightarrow \mathcal{A}_{max}$. While both $u_E$ and $|E|^4$ are significantly enhanced in the vicinity of the bridge for small $h$, this local enhancement coincides with an overall reduction in $U_{core} \mathcal{A}_{max} = \left(\frac{n_g}{n_{core}}\right) \gamma_{core}$ as observable from Figs. 3(c) and 3(d). In the v-groove geometry this results in *suppressed* non-linearity regardless of the cladding refractive index. Notably, in these devices it is possible to achieve a ~30-50x enhancement in a linear metric for a given medium, while simultaneously achieving a ~3x suppression in non-linearity from the same medium. This unique capability is unachievable in low-index contrast optics and offers an attractive design solution to scaling the efficiency of linear optical devices while suppressing non-linear performance impairments [49].

A simple explanation to this unique effect could be described as follows. In a linear device harnessing matter-light interaction, the active volume is defined by specifically engineering the geometry of the active material or region, which can be advantageously tailored on the nanoscale to be significantly smaller than the total dimensions of the waveguide core [34]. This enables high values of optical concentration $U_\mathcal{A}$ to be realized within the active area, assuming $\mathcal{A} < \mathcal{A}_{max}$. Non-linearity on the other hand, implies that interactions are 'pumped' (light-matter effect) and 'probed' (matter-light effect) across the entire non-linear medium. In the diabolo or v-groove waveguide geometries, *enhancement* in non-linearity, relative to a strip waveguide, could be achieved only when: (1) the dominant non-linear material is restricted to a small size, $\mathcal{A} < \mathcal{A}_{max}$, which is comparable to the region of enhanced $u_E$ and $|E|^4$ (e.g. localized interactions with atoms, defects, nanomaterials); and/or (2) the integrated $|E|^4$ enhancement term overcomes the $\Gamma_{core}^2$ suppression, as is apparent for the air-clad diabolo.

The high index contrast of the air clad diabolo waveguide results in a very large peak $|E|^4$ enhancement in silicon, which approaches with decreasing $h$ a theoretical enhancement factor of $\sim(n_h/n_l)^8 \approx 2 \times 10^4$ relative to a homogenous silicon strip waveguide core. Thus, despite its lower transverse confinement factor $\gamma_{core}$ and group index relative to a strip waveguide, the diabolo geometry enables significant reduction in $A_{eff}^{(NL)}$. For $h \approx 2$ nm, this corresponds to a record level fast-light non-linear silicon waveguide parameter $\gamma \approx 1.5 \times 10^6$ W$^{-1}$ km$^{-1}$. We also note the modal properties, including $A_{eff}^{(NL)}$ and thus the nonlinear parameter, are very weakly influenced by the v-groove tip's radius of curvature (see Fig. S1). This non-linearity enhancement is particularly impressive considering the width and height dimensions are unoptimized and that the non-linearity could be further enhanced in resonant [50] or slow-light configurations [46], if desired. In general however, waveguide systems exploiting highly localized non-linearities (atomic scale $\mathcal{A}$), such as those derived from silicon or germanium vacancy centers in diamond [51,52], would likely realize the most significant enhancements to non-linearities.

Fig. 4 depicts the computed optical concentration, in silicon, of air-clad diabolo and v-groove waveguides with small silicon bridge heights, $h = 2$ nm and 20 nm, benchmarked against the silicon strip waveguide of Fig. 1(a) and a hybrid dielectric-nanoplasmonic Si-Ag structure from the recent literature [24]. The optical concentration $U_\mathcal{A}$ is computed via numerical evaluation of Eq. (9), wherein the active area $\mathcal{A}$ is swept over a large range of possible shapes/sizes within the silicon cross-section. For large $\mathcal{A}$, comparable to or larger in scale than the diffraction limit $(\lambda_n/2)^2$, $U_\mathcal{A}$ cannot be enhanced through the transverse plasmonic, dielectric, or metamaterial design owing to energy conservation and mode normalization. This is observed mathematically in the $\gamma_\mathcal{A}/\mathcal{A}$ term from Eq. (9). Thus, in the regime of large $\mathcal{A}$, the only tools available to significantly enhance optical concentration are to reduce the group velocity or recirculate light [16,41,53].

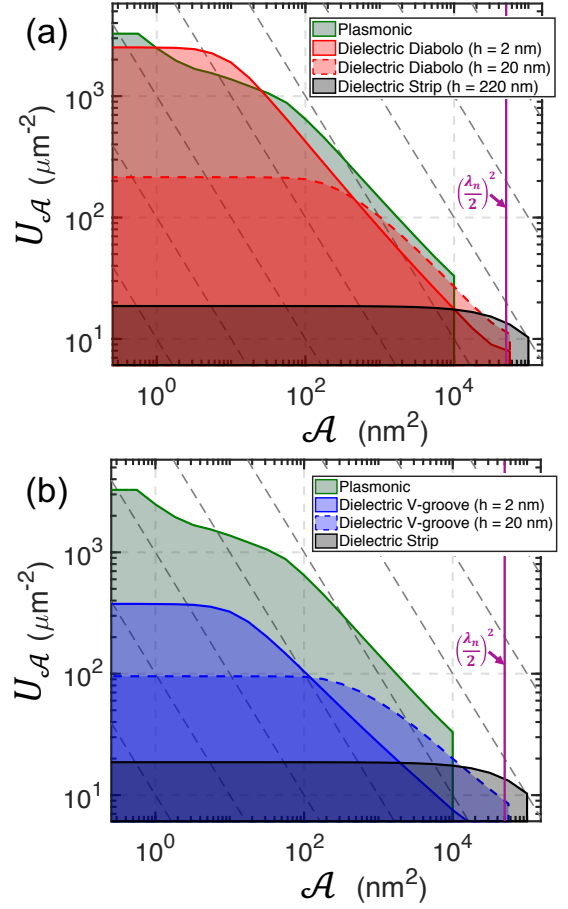

**Fig. 4.** All-dielectric sub-diffraction enhancement in optical concentration. The optical concentration (in silicon) is computed numerically via Eq. (9) vs. active area dimensions, for reference dielectric strip (500 x 220 nm) and plasmonic waveguides; and compared to air cladded silicon: (a) diabolo and (b) v-groove waveguides. Dashed diagonal lines indicate constant $U_\mathcal{A}\mathcal{A} = \Gamma_\mathcal{A}$ contours.

For small $\mathcal{A} < (\lambda_n/2)^2$ however, transverse structuring of the dielectric function, and enforcement of vectorial boundary conditions, allows the electric field energy to be significantly redistributed and locally enhanced. This enables the v-groove and diabolo waveguides to achieve extreme optical concentrations $U_\mathcal{A} \sim 10^2$ $\mu m^{-2}$ to $10^3$ $\mu m^{-2}$. Remarkably, we observe the optical concentration of the air-clad diabolo waveguide closely rivals, or in some cases even exceeds the plasmonic benchmark. Thus, while both the plasmonic and diabolo waveguides achieve a 'classic' $A_n$ on the order of $\sim\lambda_0^2/10,000$, when considering $r_{max}$ from Eq. (11) is in silicon, the plasmonic structure exhibits a theoretical propagation loss of $\sim 10^4$ dB cm$^{-1}$ while the all-dielectric diabolo is theoretically lossless.

The optical concentration profiles in Fig. 4 show clear plateaus where further reducing $\mathcal{A}$ provides diminishing and ultimately negligible improvement in $U_\mathcal{A}$. This is associated with the active dimensions becoming comparable to and ultimately smaller than the extent of the localized $u_E$ enhancement. In a strip waveguide, this plateau occurs in the vicinity of the diffraction limit, e.g. near $\mathcal{A} \approx 10^4$ nm$^2$. Such an active area could be realized with a $\sim 100$ nm x 100 nm active region, or with some slight penalty a $\sim 45$ nm x 220 nm active region. Such dimensions are comparable to the active regions of state-of-the-art *pn* diode electro-optic modulators [54]. While the active regions of such devices can be made larger to increase $\Gamma_\mathcal{A}$, i.e. using wider depletion regions or wrapped junctions [54,55], this only optimizes the device length *l*, i.e. Eq. (7), and in fact penalizes the optical concentration $U_\mathcal{A}$. Per Eq. (10), any reduction in $U_\mathcal{A}$ must be met with an increase in active volume $\mathcal{V}$ to achieve the desired optical response. Similarly, any device which can achieve a larger $U_\mathcal{A}$ enables the optical response to be achieved with a lower active volume $\mathcal{V}$. An important implication of this, is that the diabolo or v-groove waveguide, for example, could principally enable reduction in the active volume of a state-of-the-art silicon diode phase shifter by a factor of $\sim 10$-$100$. Crucially, this does not require a plasmonic structure to be utilized and can therefore potentially be realized with the low optical losses typically associated with all-dielectric media. Given that the grooves are in principle amenable to fabrication via anisotropic wet etching of silicon [43,44], we expect that very smooth surfaces can ultimately be realized – much smoother than traditional reactive ion-etched sidewalls. Thus, it's plausible to expect optical losses might be on par with the $\sim$dB cm$^{-1}$ scale losses of standard silicon nanowires.

The silicon phase shifter, however, is only one niche example of the implications of achieving an extreme optical concentration [34]. The exact same efficiency and active volume scaling principles apply to the design of waveguide integrated light emitters and absorbers. It should also be noted that in many practical cases of interest a $\sim 10$x enhancement in $U_\mathcal{A}$, for example, may require working with a $\sim 100$x reduction in $\mathcal{A}$. Perhaps counterintuitively, this configuration enables a reduction in total active volume $\mathcal{V}$ of $\sim 10$x but requires simultaneously *lengthening* the device by a factor of $\sim 10$x. This highlights the significance of the transverse device dimensions, which in fact control 2 out of 3 spatial degrees of freedom. In device applications where $\mathcal{A}$ is already very small for fundamental reasons (i.e. atom-light interfaces, integrated 2D atomic materials, quantum wells, etc.) then $\sim 10$-$100$x enhancement in $U_\mathcal{A}$ is feasible under constant $\mathcal{A}$. In such cases, the efficiency *and* length of the device can be improved by the $\sim 10$-$100$x factor relative to a diffraction limited waveguide. As a powerful example, one could already imagine a diabolo or v-groove bridge, with height *h*, being formed entirely out of a single sub-nanometer thickness high index 2D atomic monolayer. The enforcement of vectorial boundary conditions would constrain the mode solution to yield unprecedented low-loss optical concentration in the active nanomaterial. These results and observations clearly indicate that continued investigations into the regime of deeply sub-wavelength dielectric nanophotonics are warranted and will likely yield new generations of ultra-efficient linear and non-linear devices.

## V. CONCLUSION

In summary, we have introduced a simple approach for designing all-dielectric waveguides capable of achieving significantly enhanced linear and non-linear interactions. Moreover, we have laid out the theory of optical concentration in the context of waveguides and shown its convenient and unifying characteristics in describing the performance of linear and non-linear devices with arbitrary active dimensions.

The principle physics investigated here, rely on vectorial boundary conditions to Maxwell's equations. Indeed, the vector nature of light offers a powerful tool for tailoring light-matter interactions at the nanoscale, giving rise to birefringence, surface plasmon and slot waveguide field and energy density enhancements, metamaterial effects, deep sub-diffraction photonic crystal mode volume reduction, and now all-dielectric waveguide field and energy density enhancements accessible by high index media. We envision a wide array of scientific and technological applications that may benefit from the now expanded nanophotonic 'toolkit', including for example ultra-efficient integrated photonic active devices in both planar (e.g. nanowire) and arrayed (e.g. meta/nano-pillar) formats; high efficiency sources of classical and quantum light; broadband, slow-light, or resonant non-linear optical devices; and more.

## ACKNOWLEDGEMENTS

This work was supported by the U.S. Air Force Office of Scientific Research (AFOSR) under Grant No. FA9550-19-1-0057 (G. Pomrenke). The authors acknowledge fruitful discussions with M. Liscidini.

# APPENDIX

## 1. Methods

To assess the characteristics of different waveguides we compute their modal properties using a commercially available eigenmode solver (Lumerical MODE). Mode properties are then determined by numerical evaluation of the appropriate equation noted in the text. All calculations are performed at $\lambda_0 = 1550$ nm. All models assume the refractive indices of Si, SiO$_2$, and air to be 3.5, 1.444, and 1.0 respectively, while the complex relative permittivity of Ag in the plasmonic benchmark model is taken from Palik [56]. Unless otherwise noted, all calculations involving groove tips or corners are modelled with a realistic non-zero radius of curvature $r = 3$ nm and an ultra-fine local mesh size of 0.2 nm to ensure a fully converged mode solution which yields accurate and stable results. The groove angles are set to 54.7 degrees to mimic the potential shape of a wet etched {100} silicon microstructure [43,44]. This approach eliminates the non-physical singularity that would occur for $r = 0$ nm corners [5,57], which would result, for example, in a non-convergent calculation of $A_n$ with reducing mesh size. This same principle is applied to the slot waveguide mode area calculation in Fig. 1(ii), wherein the maximum energy density is taken from the middle of the structure and not the corner singularities as done in Ref. [5]. In our diabolo and v-groove structures the modal characteristics are observed to be a weak function of radius of curvature as shown in Appendix 2, Fig. S1; further confirming the generality of the approach considered here.

## 2. Modal Properties vs. Groove Radius of Curvature

Similar to Fig. 3 in the main text, here we calculate the modal properties of air and SiO$_2$ cladded diabolo and v-groove waveguides as a function of groove tip radius of curvature $r$. Here the silicon bridge height $h$ is fixed to 2 nm. The results further validate our findings in the manuscript and show the modal properties are weakly affected by groove radius.

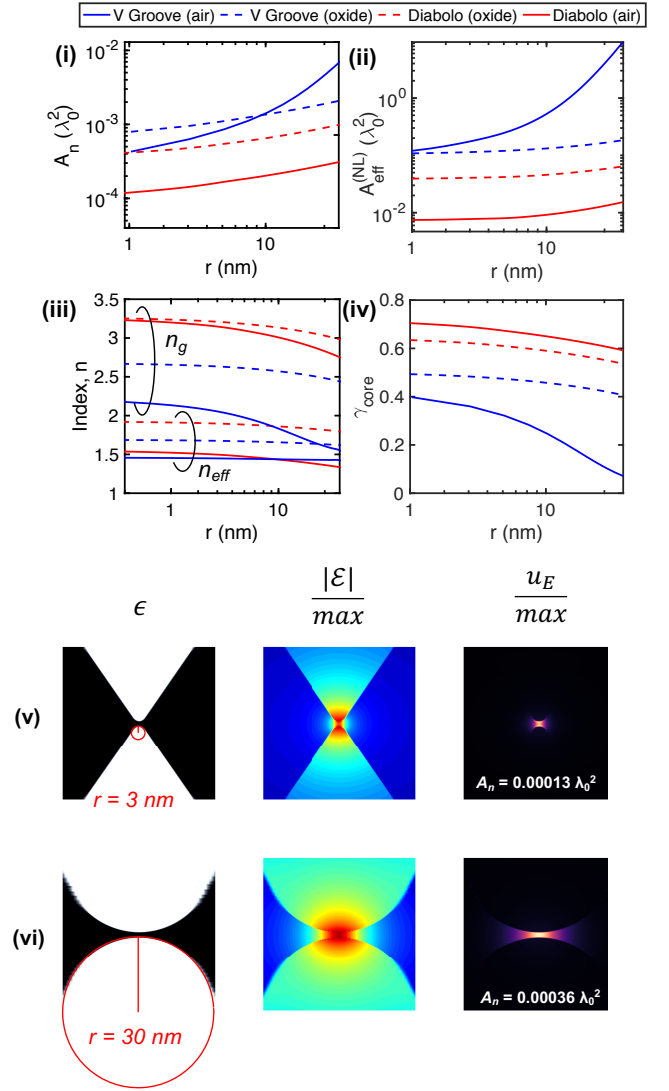

**Fig. S1.** Modal properties of v-groove and diabolo waveguides as a function of groove tip radius of curvature $r$. Here the silicon bridge height $h$ is fixed to 2 nm.

## 3. Linear Interactions and Their Relationship to "Optical Concentration"

The response of linear active waveguide devices may be described by a complex perturbation in wavevector according to [40]:

$$\Delta\beta = \Gamma_\mathcal{A}\left(\frac{\omega}{c}\Delta n_\mathcal{A} + \frac{i}{2}\alpha_\mathcal{A}\right) \\ = \Gamma_\mathcal{A}\left(\frac{\omega}{c}\Delta n_\mathcal{A} - \frac{i}{2}g_\mathcal{A}\right) \quad (A3.1)$$

Which as noted in Eq. (9) may be relayed in terms of optical concentration $U_\mathcal{A}$ per the relation:

$$U_\mathcal{A} = \Gamma_\mathcal{A}/\mathcal{A} \quad (A3.2)$$

The derivation of these metrics, $\Gamma_{\mathcal{A}}$ and $U_{\mathcal{A}}$, beginning from Eq. (5) of the main text are as follows:

$$\Delta\tilde{\beta} = \frac{\omega \int \langle \Delta u \rangle dA}{\omega \int \langle (\partial u/\partial |\boldsymbol{k}|) \cdot \hat{z} \rangle dA} \quad (A3.3)$$

Here, the denominator of Eq. (5) or (A3.3) describes the time averaged total energy flux across a plane perpendicular to the z-axis, and can be alternatively re-expressed for a travelling wave according to the relation:

$$\omega \int \langle (\partial u/\partial |\boldsymbol{k}|) \cdot \hat{z} \rangle dA = \frac{\partial \omega}{\partial k} \int 2 \langle u_E \rangle dA \quad (A3.4)$$

Which is a statement that energy flows at the group velocity $\frac{\partial \omega}{\partial k} = c/n_g$ and accounts for the total energy per unit length $\int \langle u \rangle dA$ being equal to twice the total electric field energy per unit length $2\int \langle u_E \rangle dA$. For non-magnetic optical devices ($\mu_r = 1$), local energy density perturbations are manifested strictly through material permittivity such that $\langle \Delta u \rangle = \langle \Delta u_E \rangle$. Therefore, in the approximation of low material dispersion the numerator of main text Eq. (5) can be re-expressed:

$$\omega \int \langle \Delta u_E \rangle dA = 2\omega \int \frac{\Delta \tilde{n}_{\mathcal{A}}(\boldsymbol{r})}{n_{\mathcal{A}}(\boldsymbol{r})} \langle u_E \rangle dA \quad (A3.5)$$

Where $\Delta \tilde{n}_{\mathcal{A}}$ is a complex perturbation in material refractive index and $n_{\mathcal{A}}$ is the unperturbed refractive index. For a spatially invariant $\Delta \tilde{n}_{\mathcal{A}}$, occurring only in an active region $\mathcal{A}$, the index fraction may be pulled outside the integral.

Then substituting Eqs. (A3.4) and (A3.5) into Eq. (A3.3) yields a solution consistent with Ref. [40] from the main text:

$$\Delta\tilde{\beta} = \frac{\omega}{c} \frac{n_g}{n_{\mathcal{A}}} \frac{\int_{\mathcal{A}} u_E dA}{\int u_E dA} \Delta \tilde{n}_{\mathcal{A}} \quad (A3.6)$$

Which is equivalent with or without time averaging and recovers a form of the rigorously derived confinement factor $\Gamma_{\mathcal{A}}$, which is the coefficient that is known to satisfy:

$$\Delta\tilde{\beta} = \frac{\omega}{c} \Gamma_{\mathcal{A}} \Delta \tilde{n}_{\mathcal{A}} \quad (A3.7)$$

The accumulated response over some length $l$ is therefore:

$$\Delta\tilde{\beta} l = \frac{\omega}{c} \Gamma_{\mathcal{A}} \Delta \tilde{n}_{\mathcal{A}} l = \frac{\omega}{c} \Gamma_{\mathcal{A}} \Delta \tilde{n}_{\mathcal{A}} \frac{\mathcal{V}}{\mathcal{A}} \quad (A3.8)$$

This leads to Eq. (10) of the main text:

$$\Delta\tilde{\beta} l = \frac{\omega}{c} U_{\mathcal{A}} \mathcal{V} \Delta \tilde{n}_{\mathcal{A}} \quad (A3.9)$$

Where the optical concentration is the coefficient defined in Eq. (8) of the main text which follows from Eqs. (A3.3-8):

$$U_{\mathcal{A}} \equiv \frac{\int_{\mathcal{A}} \langle u_E \rangle dA}{\mathcal{A}} \left(\frac{c}{n_{\mathcal{A}}}\right) \frac{2}{\omega \int \langle (\partial u/\partial |\boldsymbol{k}|) \cdot \hat{z} \rangle dA} \quad (A3.10)$$

And can also be expressed according to Eq. (9).

For the case of a linear waveguide with continuous translational symmetry, it is convenient to utilize $U_{\mathcal{A}}$, whereas for a periodic waveguide (i.e. subwavelength grating, metamaterial, or photonic crystal) an alternative definition may be used. A perturbation in complex wavevector arising from a *linear* interaction is therefore:

$$\Delta\tilde{\beta} = U_{\mathcal{A}} \mathcal{A} \left(\frac{\omega}{c} \Delta n_{\mathcal{A}} + \frac{i}{2} \alpha_{\mathcal{A}}\right)$$
$$= U_{\mathcal{A}} \mathcal{A} \left(\frac{\omega}{c} \Delta n_{\mathcal{A}} - \frac{i}{2} g_{\mathcal{A}}\right) \quad (A3.11)$$

The complex phase shift accumulated over some propagation length $l$ is therefore:

$$\Delta\tilde{\beta} l = U_{\mathcal{A}} \mathcal{A} l \left(\frac{\omega}{c} \Delta n_{\mathcal{A}} + \frac{i}{2} \alpha_{\mathcal{A}}\right)$$
$$= U_{\mathcal{A}} \mathcal{A} l \left(\frac{\omega}{c} \Delta n_{\mathcal{A}} - \frac{i}{2} g_{\mathcal{A}}\right) \quad (A3.12)$$

$$\Delta\tilde{\beta} l = U_{\mathcal{A}} \mathcal{V} \left(\frac{\omega}{c} \Delta n_{\mathcal{A}} + \frac{i}{2} \alpha_{\mathcal{A}}\right)$$
$$= U_{\mathcal{A}} \mathcal{V} \left(\frac{\omega}{c} \Delta n_{\mathcal{A}} - \frac{i}{2} g_{\mathcal{A}}\right) \quad (A3.13)$$

The complex phase shift, which is a determinizing characteristic of the active device response, is therefore proportional to the optical concentration and active volume:

$$\beta l_z \propto \mathcal{V} U_{\mathcal{A}} \quad (A3.14)$$

This relationship expresses a clear scaling principle applicable in general to all linear photonic waveguide based devices. It may be simply summarized as follows: for a constant stimulus (i.e. perturbation in material properties) and constant complex phase shift, minimization of the active device volume (which is principally proportional to the minimum energy consumption) requires maximization of the optical concentration $U_{\mathcal{A}}$.

## 4. Purcell Factor and Mode Area and Their Relationship to "Optical Concentration"

Taking the spontaneous emission rate $\Gamma_s$ to be proportional to optical concentration $U_\mathcal{A}$, a simple definition of the Purcell factor in a waveguide could be made using a ratio of optical concentrations as:

$$F_P = \frac{\Gamma_s}{\Gamma_0} = \frac{U_\mathcal{A}}{U_0} \quad (A4.1)$$

Which should agree with the conventionally defined Purcell factor under the appropriate reference concentration $U_0$. If considering only a single dipole at the field maximum, this relationship becomes

$$F_P = \frac{U_{\mathcal{A} \to dA}|_{r=r_{max}}}{U_0} = \frac{1}{U_0}\left(\frac{\eta}{A_n}\right) \quad (A4.2)$$

Where $A_n$ is the waveguide mode area:

$$A_n = \frac{\int u_{E_n} dA}{u_{E_n}(r_{max})} \quad (A4.3)$$

Per the approach taken by Miller, e.g. Ref. [34], it has been shown that the reference concentration $U_0$ may be described as $U_0 = \left(\frac{3}{2\pi}\right)\lambda_n^2$, where $\lambda_n = \frac{\lambda_0}{n}$ is the wavelength in the material with refractive index $n$. Substituting into Eq. (A4.3) then yields the waveguide Purcell factor:

$$F_P = \left(\frac{3}{2\pi}\right)\lambda_n^2 U_\mathcal{A} = \frac{3}{2\pi}\left(\frac{\eta}{\tilde{A}_n}\right) \quad (A4.4)$$

Where the mode area can be normalized into units of $\lambda_n^2$ according to

$$\tilde{A}_n = \frac{\int u_{E_n} dA}{u_{E_n}(\boldsymbol{r}_{max})}\left(\frac{n(\boldsymbol{r}_{max})}{\lambda}\right)^2 \quad (A4.5)$$

If the waveguide is formed into a cavity, then the concentration enhancement provided by the $\eta$ term is replaced with a factor $\mathfrak{F}/\pi$, where $\mathfrak{F}$ is the cavity finesse [34].

$$F_P = \frac{3}{2\pi^2}\left(\frac{\mathfrak{F}}{\tilde{A}_n}\right) \quad (A4.6)$$

In high-finesse cavities the finesse and Q-factor are related according to:

$$\mathfrak{F} = \frac{\lambda_n Q}{2L} \quad (A4.7)$$

Therefore, the Purcell factor becomes:

$$F_P = \frac{3}{4\pi^2}\left(\frac{Q}{\tilde{A}_n}\right)\frac{\lambda_n}{L} = \frac{3}{4\pi^2}\left(\frac{Q}{\tilde{V}_n}\right) \quad (A4.8)$$

Which recovers the classic unitless definition of the Purcell factor in terms of mode volume $\tilde{V}_n$ normalized into units of $\lambda_n^3$.

## 5. Non-linear Interactions and Their Relationship to "Optical Concentration"

Non-linear interactions, such as degenerate four-wave mixing, self-phase modulation, or two-photon absorption in bulk optical media can be described by an intensity [W/m²] dependent complex perturbation in complex wave vector:

$$\Delta k = \left(\frac{\omega}{c}\Delta n_{NL} + \frac{i}{2}\alpha_{NL}\right) \quad (A5.1)$$

Where the change in material index can be described by:

$$\Delta n_{NL} = n_2 I = n_2 \frac{P}{A} \quad (A5.2)$$

Here the underlying physics of, for example, a $\chi^{(3)}$ process and intensity dependent polarization describing the light-matter interaction, are simply captured in a macroscopic model, where $n_2$ is a material property [RIU m² / W], and $I$ is the optical intensity in the material [W / m²], which could alternatively be described by the input optical power $P$ [W] divided by an area $A$ [m²].

The non-linearly induced absorption coefficient may similarly be described by:

$$\alpha_{NL} = \beta_T I = \beta_T \frac{P}{A} \quad (A5.3)$$

Here the underlying physics (i.e. mediated by two-photon absorption) are again captured in a macroscopic model via a material coefficient $\beta_T$ [m / W]. A non-linear coefficient $\gamma$ can thus be simply defined by factoring the input power P out of the complex perturbation in complex wavevector:

$$\tilde{\gamma} \equiv \frac{\Delta k}{P} = \left(\frac{\omega}{c}\Delta n_{NL} + \frac{i}{2}\alpha_{NL}\right)/P \quad (A5.4)$$

$$\tilde{\gamma} = \frac{1}{A}\left(\frac{\omega}{c}n_2 + \frac{i}{2}\beta_T\right) \quad (A5.5)$$

In an optical waveguide the non-linear coefficient captures the complex perturbation $\Delta\tilde{\beta}$ in the waveguide's complex wavevector $\tilde{\beta}$.

$$\tilde{\gamma} = \frac{\Delta \tilde{\beta}}{P} = \left(\frac{\omega}{c}\Delta n_{eff,NL} + \frac{i}{2}\alpha_{eff,NL}\right)/P \quad (A5.6)$$

By treating the coefficients $n_2$ and $\beta_T$ as material parameters (which are valid under the general macroscopic form of Maxwell's equations) we can write the non-linear coefficient in terms of a non-linear effective mode area $A_{eff}^{(NL)}$ [m$^2$]:

$$\tilde{\gamma} = \frac{1}{A_{eff}^{(NL)}}\left(\frac{\omega}{c}n_2 + \frac{i}{2}\beta_T\right) \quad (A5.7)$$

Clearly the non-linear parameter is maximum when $A_{eff}^{(NL)}$ is minimized and vice versa. An exact and correct calculation of $A_{eff}^{(NL)}$ is therefore crucial to assist in designing optical devices to either enhance or suppress non-linear effects. One common definition often utilized in the literature is:

$$A_{eff}^{(i)} = \frac{\left(\iint_\infty |\mathcal{E}_n(\mathbf{r})|^2 dA\right)^2}{\iint_{NL} |\mathcal{E}_n(\mathbf{r})|^4 dA} \quad (A5.8)$$

However, this expression is only valid in the limit of vanishing index contrast $(n_H - n_L) \approx 0$ and vanishing contrast between material and waveguide group indices $(n_{g,NL} - n_{g,wvg}) \approx 0$, a regime which is clearly inapplicable to most integrated optical devices. A correct expression for $A_{eff}^{(NL)}$ crucially requires accounting for the fully vectorial nature of wave propagation and the power and/or energy distribution and confinement in the waveguide. Only very recently has a correct fully vectorial mode area been derived from Maxwell's equations and supported experimentally, see Ref. [47]:

$$A_{eff}^{(f)} = \frac{3\left(\iint_\infty n(\mathbf{r})^2 |\mathcal{E}|^2 dA\right)^2}{n_{g,wvg}^2 n_H^2 \iint_{NL:n=n_H} \mathcal{E}^* \cdot [2|\mathcal{E}|^2 \mathcal{E} + (\mathcal{E}\cdot\mathcal{E})\mathcal{E}^*] dA} \quad (A5.9)$$

However, the calculation in this exact form is rather cumbersome and does not provide clear linkage to other common metrics such as confinement factor or our metric of optical concentration. Here we independently derive an alternative expression for $A_{eff}^{(NL)}$, which fully considers the aforementioned criteria, and present it in an easily calculable form accessible to most researchers. The details/benefits of this approach will be further examined in a forthcoming manuscript, however, are included here for review purposes and completeness. Further, we identify for the first time, a clear relationship between the exact vectorial non-linear mode area and a rigorous definition of the optical confinement factor used in linear optics. This suggests the non-linear parameter and mode area can be determined from experimental measurement of the *linear* confinement factor paired with a calculable *non-linear* correction factor.

Our derivation of $A_{eff}^{(NL)}$ relies on the macroscopic form of Maxwell's equations describing the fully vectorial nature of electromagnetic propagation and confinement and takes the assumption of single-mode degenerate four wave mixing (FWM) such that a single field profile may be considered.

Recall from the above section (Appendix 3) regarding linear matter-light interactions, inducing a complex perturbation $\Delta\beta$ to the waveguide wavevector $\beta$ according to:

$$\begin{aligned}\Delta\tilde{\beta} &= \Gamma_\mathcal{A}\left(\frac{\omega}{c}\Delta n_\mathcal{A} + \frac{i}{2}\alpha_\mathcal{A}\right)\\ &= \Gamma_\mathcal{A}\left(\frac{\omega}{c}\Delta n_\mathcal{A} - \frac{i}{2}g_\mathcal{A}\right)\end{aligned} \quad (A5.10)$$

In this form $\Delta n_\mathcal{A}$ and $\alpha_\mathcal{A}$ are treated as perturbations of the unperturbed material properties such that its refractive index is perturbed uniformly within the active region $\mathcal{A}$ according to $\tilde{n} = n + \Delta\tilde{n}$, where $\Delta\tilde{n} = \Delta n_\mathcal{A} + i\frac{4\pi}{\lambda}\alpha_\mathcal{A} = \Delta n_\mathcal{A} - i\frac{4\pi}{\lambda}g_\mathcal{A}$. The confinement factor $\Gamma_\mathcal{A}$ is therefore the coefficient which satisfies:

$$\Delta\tilde{\beta} = \frac{\omega}{c}\Delta\tilde{n}\Gamma_\mathcal{A} = \frac{\omega}{c}\Delta\tilde{n}_{eff} \quad (A5.11)$$

$$\Delta\tilde{n}_{eff} = \Delta\tilde{n}\Gamma_\mathcal{A} \quad (A5.12)$$

Note: The confinement factor often appears in the literature incorrectly as a measure of fraction of total electromagnetic power propagating along the z-axis confined in the active region (using either integrated Poynting vector or field intensity) normalized to the total electromagnetic power propagating along the z-axis. Such expressions are only valid only in the limit of vanishing index contrast $(n_H - n_L) \approx 0$ and vanishing contrast between material and waveguide group indicies $(n_{g,NL} - n_{g,wvg}) \approx 0$. The correct form of the confinement factor, which has been derived via the variational principle, e.g. Eq. (A3.6) and Ref. [40], and has been shown to capture the physics of fully vectorial fields and modal dispersion in high index contrast media can also be written as:

$$\Gamma_\mathcal{A} = \frac{n_\mathcal{A} c \epsilon_0 \iint_\mathcal{A} |\mathcal{E}|^2 dA}{\iint_\infty Re\{\mathcal{E}\times\mathcal{H}^*\}\cdot\hat{\mathbf{z}} dA} \quad (A5.13)$$

Which is a measure of field intensity confined to the active region normalized to unit power. Although it may not appear obvious, this expression does in fact capture the effect of the waveguide group index $n_g$,

$$n_g = \frac{c \iint_\infty \frac{1}{2}\frac{\partial}{\partial \omega}(\omega \epsilon)|\mathcal{E}|^2 dA}{\frac{1}{2}\iint_\infty Re\{\mathcal{E} \times \mathcal{H}^*\} \cdot \hat{z} dA} \quad (A5.14)$$

which can rigorously be calculated even from a single frequency mode calculation if the frequency dependence of the permittivity is term is included in the calculation, or more readily if material dispersion is small such that $\frac{\partial}{\partial \omega}(\omega \epsilon) = \epsilon$. In such a case $\Gamma_\mathcal{A}$ can be written as it is in the main text:

$$\Gamma_\mathcal{A} = \frac{n_g \iint_\mathcal{A} u_E dA}{n_A \iint_\infty u_E dA} \quad (A5.15)$$

Use of the confinement factor, however, assumes that the perturbation is strictly uniform through the active region. In a non-linear interaction, the local index change is proportional to the local electric field intensity $|E|^2$. Thus $\Delta \tilde{n}_\mathcal{A}$ is not uniform or constant across the active region and instead should be kept inside the integral, unlike Eq. 8 of the main text, such that we write:

$$\Delta \tilde{\beta} = \frac{\omega}{c}\frac{n_g}{n_\mathcal{A}}\frac{\int \Delta \tilde{n}_\mathcal{A}(x,y) u_E dA}{\int u_E dA} \quad (A5.16)$$

Or equivalently:

$$\Delta \tilde{\beta} = \frac{\omega}{c}\frac{n_\mathcal{A} c \epsilon_0 \iint_\mathcal{A} \Delta \tilde{n}_\mathcal{A}(x,y)|\mathcal{E}|^2 dA}{\iint_\infty Re\{\mathcal{E} \times \mathcal{H}^*\} \cdot \hat{z} dA} \quad (A5.17)$$

This expression assumes only that the unperturbed refractive index $n_\mathcal{A}$ is uniform within the active region. The complex index change within the waveguide cross-section can be written as:

$$\Delta \tilde{n}_\mathcal{A}(x,y) = \left(n_2 + i\frac{4\pi}{\lambda_0}\beta_T\right) I(x,y) \quad (A5.18)$$

Where the local electric field intensity function $I(x,y)$ [W/m$^2$] can be expressed in terms of the input power $P$ according to:

$$I(x,y) = \frac{n_\mathcal{A} c \epsilon_0 |\mathcal{E}(x,y)|^2}{\iint_\infty Re\{\mathcal{E} \times \mathcal{H}^*\} \cdot \hat{z} dA} P \quad (A5.19)$$

The complex change in wave-vector therefore captures both the light-matter interaction which induces the complex change in material refractive indices and the matter-light interaction which translates the change in material indices into a complex change in propagation constant, and can be expressed as either:

$$\Delta \tilde{\beta} = \frac{\omega}{c}\left(n_2 + i\frac{4\pi}{\lambda_0}\beta_T\right)\left[\frac{n_g}{n_\mathcal{A}}\frac{\int \frac{n_\mathcal{A} c \epsilon_0 |\mathcal{E}(x,y)|^2}{\iint_\infty Re\{\mathcal{E} \times \mathcal{H}^*\} \cdot \hat{z} dA} u_E dA}{\int u_E dA}\right] P \quad (A5.20)$$

Or equivalently,

$$\Delta \tilde{\beta} = \frac{\omega}{c}\left(n_2 + i\frac{4\pi}{\lambda_0}\beta_T\right)\left[\frac{\iint_\mathcal{A}\frac{(n_\mathcal{A} c \epsilon_0)^2 |\mathcal{E}(x,y)|^4}{\iint_\infty Re\{\mathcal{E} \times \mathcal{H}^*\} \cdot \hat{z} dA} dA}{\iint_\infty Re\{\mathcal{E} \times \mathcal{H}^*\} \cdot \hat{z} dA}\right] P \quad (A5.21)$$

Pulling all spatially invariant terms out of the integral, and dividing by $P$ we may write:

$$\frac{\Delta \tilde{\beta}}{P} = \tilde{\gamma} = \frac{\omega}{c}\left(n_2 + i\frac{4\pi}{\lambda_0}\beta_T\right)\left[(n_\mathcal{A} c \epsilon_0)^2 \frac{\int |\mathcal{E}(x,y)|^4 dA}{\left(\iint_\infty Re\{\mathcal{E} \times \mathcal{H}^*\} \cdot \hat{z} dA\right)^2}\right] \quad (A5.22)$$

Given that the non-linear parameter is written in terms of the effective mode area via:

$$\frac{\Delta \tilde{\beta}}{P} = \tilde{\gamma} = \frac{\omega}{c}\left(n_2 + i\frac{4\pi}{\lambda_0}\beta_T\right)\frac{1}{A_{eff}^{(NL)}} \quad (A5.23)$$

The non-linear effective mode area can be found to be:

$$A_{eff}^{(NL)} = \frac{1}{n_\mathcal{A}^2 c^2 \epsilon_0^2}\frac{\left(\iint_\infty Re\{\mathcal{E} \times \mathcal{H}^*\} \cdot \hat{z} dA\right)^2}{\int |\mathcal{E}(x,y)|^4 dA} \quad (A5.24)$$

Where the numerator can alternatively be expressed as:

$$\left(\iint_\infty Re\{\mathcal{E} \times \mathcal{H}^*\} \cdot \hat{z} dA\right)^2 = \left(\frac{c}{n_g}\iint_\infty \frac{\partial}{\partial \omega}(\omega \epsilon)|\mathcal{E}|^2 dA\right)^2 \quad (A5.25)$$

Which yields:

$$A_{eff}^{(NL)} = \frac{1}{n_\mathcal{A}^2 n_g^2 \epsilon_0^2} \frac{\left(\iint_\infty \frac{\partial}{\partial \omega}(\omega \epsilon)|\mathcal{E}|^2 dA\right)^2}{\int |\mathcal{E}(x,y)|^4 dA} \qquad (A5.26)$$

Which in the approximation of low material dispersion, can be simplified to:

$$A_{eff}^{(NL)} = \frac{1}{n_\mathcal{A}^2 n_g^2} \frac{\left(\iint_\infty n^2 |\mathcal{E}|^2 dA\right)^2}{\iint_\mathcal{A} |\mathcal{E}|^4 dA} \qquad (A5.27)$$

This expression agrees with the non-linear effective mode area derived for photonic crystal fibers [48] and is much simpler than Eq. (A5.9). Upon close inspection of Eqs. (A5.15) and (A5.27), the above expression if found to be equivalent to:

$$A_{eff}^{(NL)} = \frac{1}{\Gamma_\mathcal{A}^2} \frac{\left(\iint_\mathcal{A} |\mathcal{E}|^2 dA\right)^2}{\iint_\mathcal{A} |\mathcal{E}|^4 dA}$$
$$= \frac{1}{(U_\mathcal{A}\mathcal{A})^2} \frac{\left(\iint_\mathcal{A} |\mathcal{E}|^2 dA\right)^2}{\iint_\mathcal{A} |\mathcal{E}|^4 dA} \qquad (A5.28)$$

Which is the formula presented in the main text, Eq. (14), This form of the non-linear effective mode area shows that the non-linearity can be explicitly linked to the confinement factor utilized in *linear* optics. The right most term is effectively a field correction term which ensures integration is performed over the $|\mathcal{E}|^4$ profile rather than the $|\mathcal{E}|^2$ profile. Given that $\Gamma_\mathcal{A}^2 = (U_\mathcal{A}\mathcal{A})^2$ the non-linear effective mode area can also be directly linked to the optical concentration.

With Eq. (A5.28) in hand we can rewrite Eq. (A5.23) directly in terms of optical concentration and active area:

$$\frac{\Delta \tilde{\beta}}{P} = \tilde{\gamma} = \frac{\omega}{c}\left(n_2 + i\frac{4\pi}{\lambda_0}\beta_T\right) \frac{(U_\mathcal{A}\mathcal{A})^2 \iint_\mathcal{A} |\mathcal{E}|^4 dA}{\left(\iint_\mathcal{A} |\mathcal{E}|^2 dA\right)^2} \qquad (A5.29)$$